\documentclass[conference]{IEEEtran}
\IEEEoverridecommandlockouts
\usepackage{enumitem}
\usepackage{mwe} 
\usepackage{graphicx}
\usepackage{float}
\usepackage[english]{babel}
\usepackage[utf8]{inputenc}
\usepackage{array, multirow, makecell}
\usepackage[table]{xcolor}
\usepackage{hyperref}
\usepackage{etoolbox}
\usepackage{xspace}
\usepackage[usestackEOL]{stackengine}
\usepackage{cite}
\usepackage{amsmath,amssymb,amsfonts}
\usepackage{algorithmic}
\usepackage{xurl}
\usepackage{textcomp}
\usepackage{xcolor}
\usepackage[fontsize=10pt]{fontsize}
\def\BibTeX{{\rm B\kern-.05em{\sc i\kern-.025em b}\kern-.08em T\kern-.1667em\lower.7ex\hbox{E}\kern-.125emX}}
\newcommand{\STAB}[1]{\begin{tabular}{@{}c@{}}#1\end{tabular}}
\newcommand{\etal}{\textit{et al.}}
    
\begin{document}

\title{Detection and Localization of Melanoma Skin Cancer in Histopathological Whole Slide Images}


\author{\IEEEauthorblockN{Neel Kanwal\IEEEauthorrefmark{1}$^1$,
Roger Amundsen\IEEEauthorrefmark{1}$^1$,
Helga Hardardottir\IEEEauthorrefmark{2}\IEEEauthorrefmark{3}, Luca Tomasetti\IEEEauthorrefmark{1}, Erling Sandøy Undersrud\IEEEauthorrefmark{2}\IEEEauthorrefmark{3}\\
Emiel A.M. Janssen\IEEEauthorrefmark{2}\IEEEauthorrefmark{3},  
Kjersti Engan\IEEEauthorrefmark{1}}

\IEEEauthorblockA{\IEEEauthorrefmark{1}Department of Electrical Engineering and Computer Science, University of Stavanger, Stavanger, Norway}
\IEEEauthorblockA{\IEEEauthorrefmark{2}Department of Chemistry, Bioscience and Environmental Engineering, University of Stavanger, Stavanger, Norway}
\IEEEauthorblockA{\IEEEauthorrefmark{3}Department of Pathology, Stavanger University Hospital, Stavanger, Norway}

\thanks{Corresponding author: neel.kanwal@uis.no.\newline \hspace{1em}  $^1$ These authors contributed equally.}}


%
\maketitle
\begin{abstract}
Melanoma diagnosed and treated in its early stages can increase the survival rate. A projected increase in skin cancer incidents and a shortage of dermatopathologists have emphasized the need for computational pathology (CPATH) systems. CPATH systems with deep learning (DL) models have the potential to identify the presence of melanoma by exploiting underlying morphological and cellular features. 
This paper proposes a DL method to detect melanoma and distinguish between normal skin and benign/malignant melanocytic lesions in whole slide images (WSI). Our method detects lesions with high accuracy and localizes them on a WSI to identify potential regions of interest for pathologists. The proposed method relies on using a single convolutional neural network to create localization maps first and use them to perform slide-level predictions to determine patients who have melanoma.
Our best model provides favorable patch-wise classification results with a 0.992 F1 score and 0.99 sensitivity on unseen data. The source code is publicly available at \href{https://github.com/RogerAmundsen/Melanoma-Diagnosis-and-Localization-from-Whole-Slide-Images-using-Convolutional-Neural-Networks}{Github}.
\end{abstract}

%
\section{Introduction}
\label{sec:intro}
Malignant melanoma is an aggressive type of skin cancer~\cite{chen2006diagnosing}. Though melanoma only accounts for roughly 1\% of skin cancer cases, it is the leading cause of mortality~\cite{cancerstats}. 
Melanoma skin cancer develops when melanocytes start to proliferate quickly and uncontrolled in the epidermis and dermis layers of the skin and form malignant lesions~\cite{cancerstats}. 
If not treated early, the tumor is likely to progress to another stage and can spread to other parts of the body~\cite{chen2006diagnosing}. Therefore, detection and accurate diagnosis at an early stage are of the utmost importance. 
Histopathological examination is a common practice for diagnosing cancer, where a skin sample is extracted by punch biopsy and processed to prepare a glass slide. Pathologists later use the glass slide to conduct microscopic inspection~\cite{kanwal2022}. It is often time-consuming and challenging for humans to distinguish between early-stage cancer and benign lesions. A digitized version of a glass slide, known as a whole slide image, can overcome this hurdle of traditional histopathology by involving computational analysis~\cite{tabatabaei2022residual, fuster2022, kanwal2023}. Computational pathology (CPATH) systems using deep learning (DL) techniques can automate various diagnostic tasks by learning morphological and cellular patterns and providing predictions~\cite{abels2019, kanwal2023, wahab2022, devil}. CPATH systems have a high potential to identify patients with melanoma and aid pathologists by providing a second opinion and localizing the regions of interest (ROIs). 

\begin{figure*}[ht!]
    \centering
    \includegraphics[width=18cm]{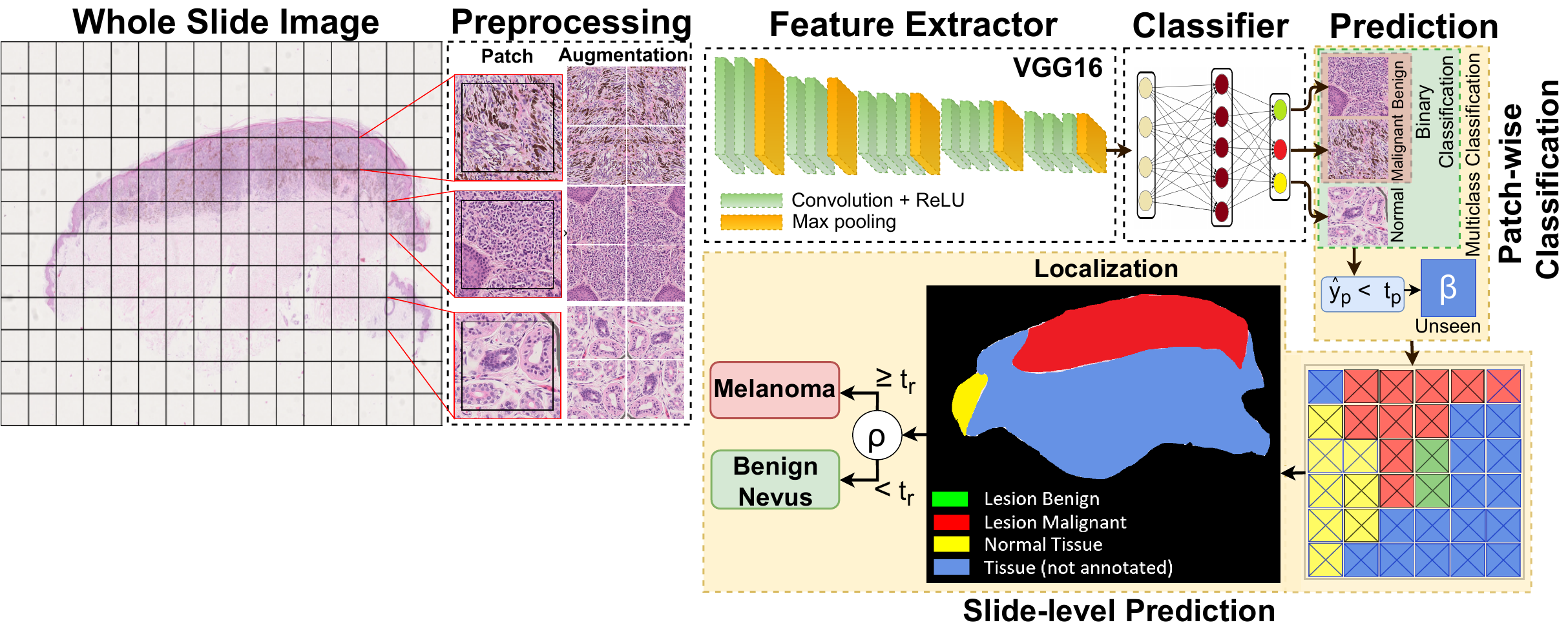}
    \caption{\textbf{An overview of our proposed melanoma detection method:} Whole slide images (WSIs) are divided into patches (sub-images) based on annotated regions. Preprocessing is performed to augment and normalize patches before feeding them to the feature extractor. VGG16~\cite{vgg} is the backbone of our model, followed by the three-layer fully-connected classifier. Binary and multiclass models perform patch-wise classification and assign a class to each patch. Probability thresholding is applied to find patches of unseen tissue and produce output with one more (unseen) class. The classified patches are then combined using their coordinates to localize lesions. Finally, the localization maps are used for slide-level predictions of \emph{melanoma} vs. \emph{benign nevus} patients. }
    \label{fig:main}
\end{figure*}

Some previous works~\cite{sheha2012automatic, jafari2016automatic} on melanoma diagnosis focused on selecting hand-crafted features from data such as first-order texture features, color intensity, entropy, eccentricity, etc. Jafari \etal~\cite{jafari2016automatic} used dermoscopic images and obtained color-space transformation to extract morphological features. Their method used the ABCD rule (Asymmetry, Border irregularity, Color patterns, and Diameter) to distinguish between malignant and benign lesions. In an analogous approach, Sheha \etal~\cite{sheha2012automatic} used gray-level co-occurrence matrix (GLCM) and texture features to discriminate between malignant melanoma and melanocytic nevi using a multilayer perceptron classifier. However, manual feature engineering is cumbersome, and data-driven features have been demonstrated to surpass classification performance in medical images~\cite{morales2021, tomasetti2021machine}. Although DL approaches automatically discover features from data, they require a significant amount of data and clinical labels, which might not always be available.

Transfer learning has been a common initialization strategy in DL to compensate for the lack of training data~\cite{kanwal2022, rajitha2022classification}. It mainly relies on transferring knowledge to a DL model, trained on a different task, for performing a new classification task by automatically learning hidden features from the new data.
Recently, Rajitha \etal~\cite{rajitha2022classification} applied transfer learning over four convolutional neural networks (CNNs) to extract features and classify dermatophytosis in unstained skin samples. Their transfer learning method focused on freezing various fractions of feature extractors on a trial-and-error basis to find the best-performing architecture. 
In a similar CNN-based approach, Logu \etal~\cite{de2020recognition} used histological images to differentiate between patches of cutaneous melanoma and healthy tissue. Their model was trained for a binary task and did not discriminate \emph{benign lesions} from malignant lesions. Zhang \etal~\cite{zhang} extracted features at multiple scales and fused them to obtain a feature representation. They performed binary patch-wise classification but did not predict melanoma at the slide level.
Wang \etal~\cite{wang2020automated} detected malignant and non-malignant nevi by CNNs and compared various architectures. Their experiments found VGG16~\cite{vgg} superior in classification performance among other CNN architectures; However, their method relied on using a separate random forest classifier to perform slide-level predictions. 



This paper proposes a CNN-based method to detect both \emph{malignant} and \emph{benign} lesions from \emph{normal} and \emph{unseen} tissue, as shown in Fig.~\ref{fig:main}, i.e., it provides a three-class prediction \emph{(four-class output)} for each patch. The classified patches are then combined into segmentation maps to localize lesions in the WSI and reveal potential ROIs for a pathologist.  Finally, the segmentation maps are used, with one extra learned parameter, to perform accurate slide-level classification as benign or malignant. Our method uses a single CNN network to perform both classification and segmentation tasks in one run.

\section{Data Materials}
\begin{figure}[t]
    \centering
    \includegraphics[width=9.5cm]{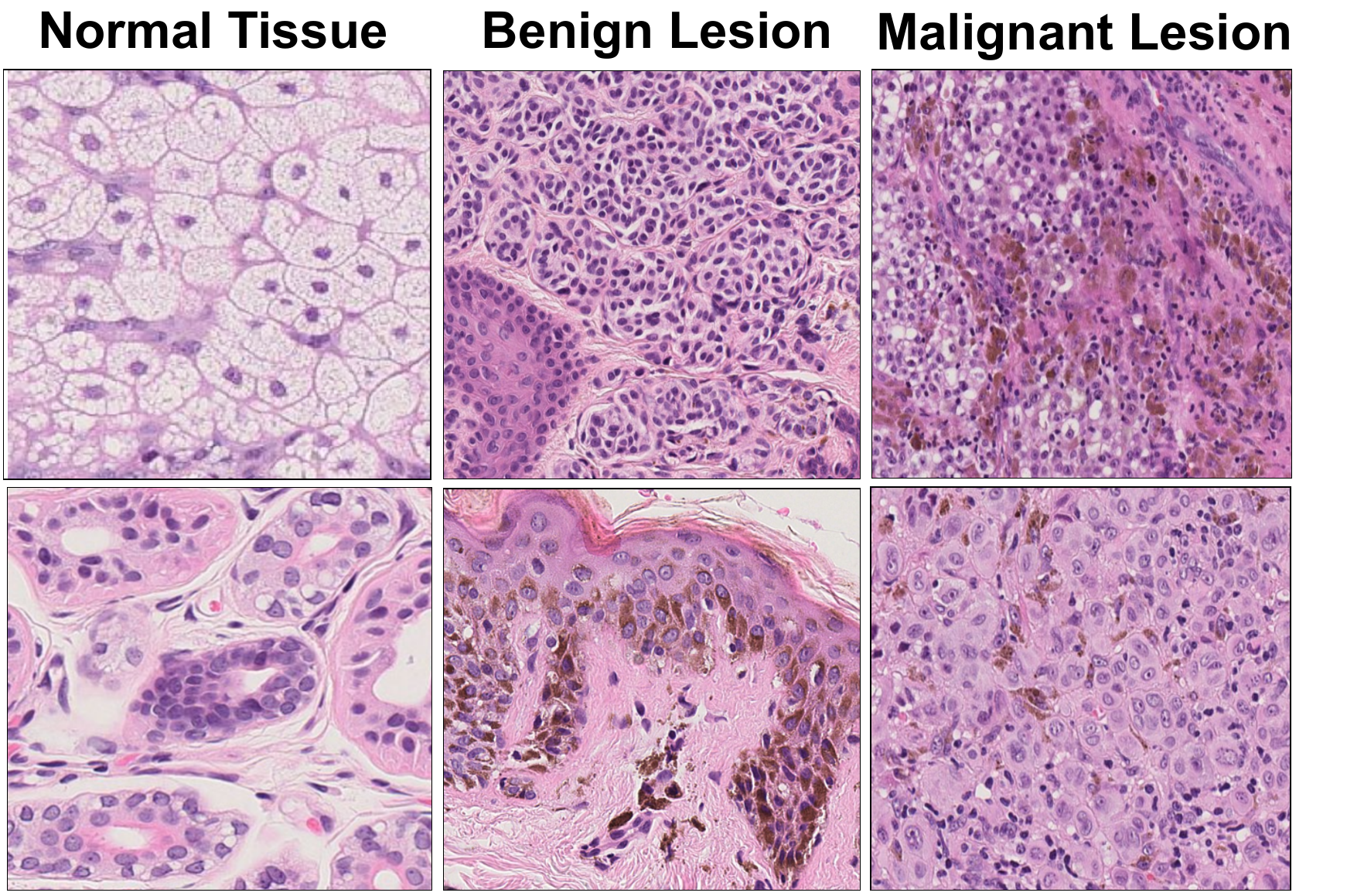}
    \caption{Sample patches of normal tissue, benign, and malignant lesion class from the dataset, extracted at 10x magnification.} 
    \label{fig:dataset}
\end{figure}

We have analyzed 90 Hematoxylin and Eosin (H\&E) stained skin biopsy whole slide images (WSIs) from Stavanger University Hospital (SUH) in Norway. These samples were clear benign or malignant, and no atypical nevi or in situ melanomas were included. The glass slides were scanned at 40x with Hamamatsu Nanozoomer s60 in \emph{ndpi} format with a pixel size of 0.2199~{\textmu}m x 0.2199~{\textmu}m. All WSIs were provided with slide-level labels of \emph{benign nevus}, \emph{benign lentigo}, and \emph{malignant melanoma} for 40, 7, and 43 WSIs, respectively. 
A split of 73/8/9 at the WSI level was carried out for training, validation, and testing. \emph{Benign nevus} and \emph{benign lentigo} were treated as a single label. A pathologist roughly annotated regions in each slide for benign lesions, malignant lesions, and other tissue types. By roughly, we mean that the annotator should limit the time spent on each WSI. Annotated regions in WSIs are divided into small sub-images (patches) for a tractable computation~\cite{kanwal2022,devil, morales2021}. 
 Fig.~\ref{fig:dataset} presents a few sample patches from each class extracted at 10x magnification level.

\section{Proposed Method}
\label{sec:method}
A graphical overview of the melanoma detection method is presented in Fig.~\ref{fig:main}. 
Pretrained VGG16~\cite{vgg} has been a popular choice for feature extractor for transfer learning in the literature~\cite{rajitha2022classification, tomasetti2022multi, wang2020automated}, and our proposed method uses it as a backbone. A custom classifier with three fully connected (FC) layers replaces the FC layers of VGG16. We proceed with the melanoma detection task by following a two-step approach. In the first step, we develop models for patch-wise classification in binary and multiclass fashion. The binary classification task distinguishes between malignant and benign lesion classes. The multiclass classification task uses three classes involving the normal tissue as well. In the second step, patches classified with a higher probability in a class are used to create a localization map and determine the slide-level outcome, as detailed below.  

\subsection{Pre-processing}
Since the CNN can not handle the entire WSI at once due to the memory, the annotated regions in WSIs were split into small patches of $256 \times 256$ pixels at 2.5x, 10x, and 40x magnification levels. First, the background-foreground segmentation was performed by transforming RGB to HSV and thresholding Hue channel $\in[100-180]$ for purple and pink tones. Later, patches with a 70\% overlap between foreground and annotation masks were extracted. We used the "patch-on-fly" for the memory-efficient patching process~\cite{devil, wetteland2021}. To even-out class imbalance, geometric transformations with random crops of $224 \times 224$ were applied to the underrepresented class. Finally, all patches were resized to $224 \times 224$ pixels to match the input size of the pre-trained model before being normalized to the mean and standard deviation of the training set.  

\subsection{Patch-wise Classification}
In the patch-wise classification step, we develop models in a binary and multiclass fashion that assign a single class to a patch. We trained a binary baseline model (Model$_{baseline}$) using a frozen feature extractor. Model$_{baseline}$ was trained on the dataset extracted at 10x magnification. 
Three binary models (Model$_{2.5x}$,  Model$_{10x}$, and Model$_{40x}$) are trained with unfrozen feature extractors using dataset extracted at 2.5x, 10x, and 40x, respectively. Our multiclass model (Model$_{multiclass}$), with three output nodes, is trained on a 10x magnification dataset. For an input patch ({$\boldsymbol{x}$) with the ground label ($y_{x}$), models output probability vector ($\boldsymbol{y}_{p}$) as shown in the Eq.~\eqref{probability}, where $\hat{y}_B$, $\hat{y}_M$, and $\hat{y}_N$ are predicted probabilities for benign lesion, malignant lesion, and normal tissue class respectively.      

\begin{equation}
 \centering
    \boldsymbol{y}_{p}  = 
  \begin{cases}
    [\hat{y}_{B}, \hat{y}_{M}] &  \quad \textrm{if}  \quad \textrm{Binary } \\
    [\hat{y}_{B}, \hat{y}_{M}, \hat{y}_{N}]  &  \quad  \textrm{if}  \quad \textrm{Multiclass}
    \end{cases}
\label{probability}
\end{equation}

Since we use the entire WSI for the slide-level prediction step, the inference stage will encounter previously unseen tissue types. To handle the unseen tissue types, a new class ($\beta$) and probability threshold (${t}_{p}$) are introduced. 
If the predicted output ($\boldsymbol{y}_{p}$) is less than ${t}_{p}$, then the predicted class ($\hat{y}_{p}$) is considered irrelevant tissue and assigned the $\beta$ label as shown in Eq.~\eqref{eq:1}. In short, the model output ($\boldsymbol{y}_{p}$) is either binary or with three classes, but the overall prediction ($\hat{y}_{p}$) includes one more class. Here, ${t}_{p}$ helps to efficiently determine any lesion's presence without training models with new tissue classes. We determine the best ${t}_{p}$ by maximizing lesion detection and minimizing false positives on the validation set. 
\begin{equation}
 \centering
    \hat{y}_{p}  = 
    \begin{cases}
    argmax(\boldsymbol{y}_{p}) &  \quad \textrm{if} \quad max(\boldsymbol{y}_{p})\geq {t}_{p} \\
                   \beta, &  \quad \textrm{Otherwise}
    \end{cases}
\label{eq:1}
\end{equation}
 
\subsection{Slide-level Prediction:}
In the slide-level prediction step, we utilize the best-performing binary and multiclass models to create masks by putting together all classified patches from the previous step. Coordinates corresponding to the patch's location are used to fill the pixel in the down-sampled map with a color. 
Later, we calculate the ratio ($\rho$) of the malignant lesion class from the localization map using Eq.~\eqref{eq:2}. We determine slide-level prediction ($\hat{y}_{s}$) of melanoma if $\rho$ is greater than a ratio (${t}_{r}$) (see Eq.~\eqref{eq:3}). In other words, ${t}_{r}$ determines the degree of malignancy to report a patient with melanoma. A suitable value of ${t}_{r}$ is selected based on the validation set to reduce the chance of false negatives, such as the classification of melanoma as a benign nevus.

\begin{equation}
\centering
    \rho  =  \frac{\text{No. of malignant pixels}}{\text{No. of malignant pixels + No. of benign pixels}}
\label{eq:2}
\end{equation}
\begin{equation}
\centering
    \hat{y}_{s}  = 
    \begin{cases}
     \textrm{Malanoma} &  \quad \textrm{if} \quad \rho \geq {t}_{r} \\
    \textrm{Benign Nevus}, &  \quad \textrm{Otherwise}
    \end{cases}
\label{eq:3}
\end{equation}

\begin{figure*}[htb]
    \centering
    \includegraphics[width=18.3cm]{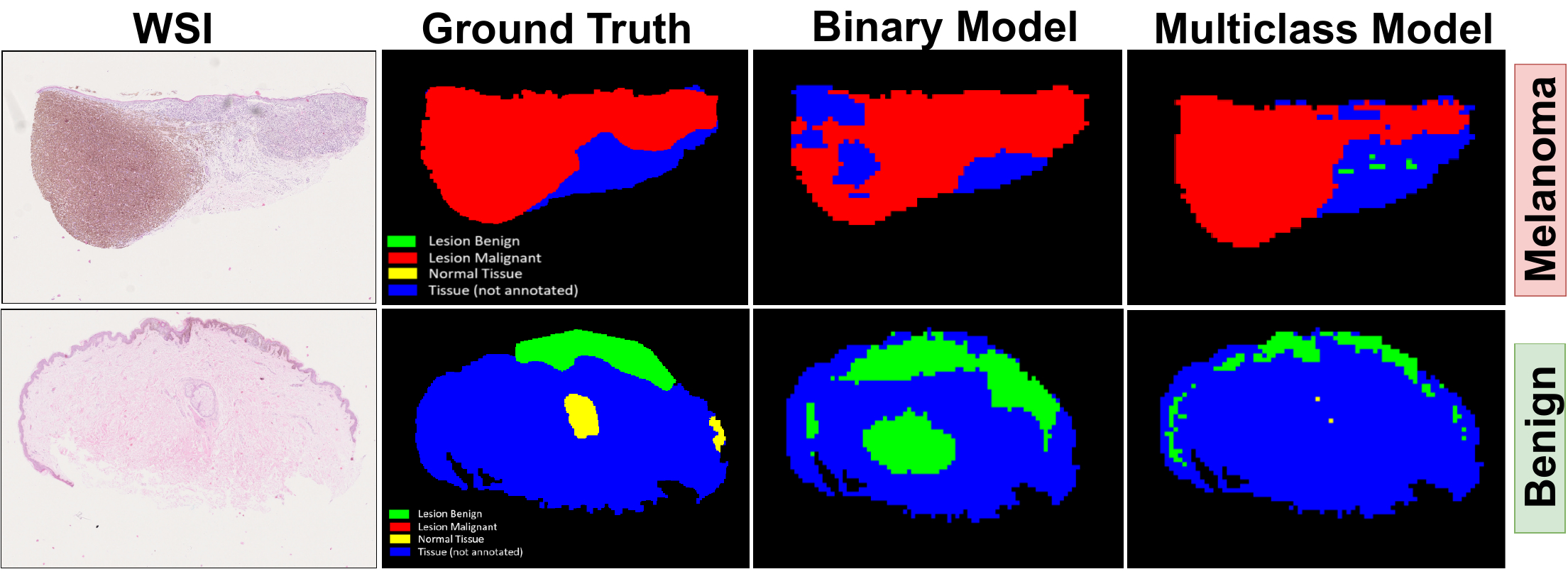}
    \caption{\textbf{Localization map for whole slide images (WSIs):} The first column represents the original WSI. The second column displays our ground truth (annotations by a pathologist). The third column shows the outcome of the binary model, and finally, the fourth column shows the outcome of the multiclass model. All WSIs, in the test set, with melanoma are correctly detected and localized.} 
    \label{fig:local}
\end{figure*}

\subsection{Evaluation Metrics}
We report accuracy, F1, sensitivity, and specificity metrics. Let TP, FN, FP, and TN describe true positive, false negative, false positive, and true negative predictions, respectively. 
Accuracy, $(TP+TN)/(TP+FN+FP+TN)$, is the percentage of accurate predictions made by the model.
$F1 = 2\cdot(\textrm{precision} \cdot \textrm{recall})(\textrm{precision} + \textrm{recall})$, is harmonic mean where Recall = Sensitivity = $TP/(TP+FN)$ and Precision = $TP/(TP+FP)$.  
The sensitivity measures patches detected as malignant lesions were actual malignant class. The specificity, termed as $TN/(TN+FP)$, describes the proportion of correctly classified benign lesion patches. 

\subsection{Implementation Details}
The method was implemented on the Pytorch~\footnote{https://github.com/pytorch/pytorch} DL framework, and patch extraction was accomplished using Pyvips~\footnote{https://libvips.github.io/pyvips/} library. VGG16~\cite{vgg} feature extractor was initialized with ImageNet weights, and the classifier was initialized with random weights. We used cross-entropy loss, Adam optimizer, the learning rate of 0.0003, batch size of 256, and early-stopping of 8 epochs on validation accuracy.  All models were developed on a Nvidia A100 40GB GPU. Source code is available at \href{https://github.com/RogerAmundsen/Melanoma-Diagnosis-and-Localization-from-Whole-Slide-Images-using-Convolutional-Neural-Networks}{Github}.

\begin{table}[hbt!]
    \centering
      \caption{The results from the patch-wise classification on the validation and test set.}
    \renewcommand{\arraystretch}{1.1}
    \resizebox{\columnwidth}{!}{\begin{tabular}{||c|l| c|c|c|c||}
        \hline
        \multicolumn{2}{||c|}{\textbf{Models}} &  \textbf{Acc. (\%)} & \textbf{F1} & \textbf{Sens.} & \textbf{Spec.}   \\ [0.2em]
        \hline 

         {} & Logu  \etal~\cite{de2020recognition} & 96.5 & 0.965 & 0.957 & 0.977 \\
         {} & Zhang \etal~\cite{zhang} & 95.5 & 0.976 & 0.986 & {-} \\
           {} &  Wang  \etal~\cite{wang2020automated} & 94.9 & {-} & 0.947 & 0.953 \\
         {} & Model$_{baseline}$ & 93.1 & 0.948 & 0.926  &  0.942  \\ 
        \cline{2-6}
        {} & Model$_{2.5x}$  & \textbf{99.32} & \textbf{0.995} & 0.990  & \textbf{0.999}   \\ 
        {} & Model$_{10x}$  & 98.13 & 0.986 &  \textbf{0.994} &  0.954 \\
        {}  & {Model$_{40x}$}  & 96.02 & 0.971 &  0.986  & 0.906 \\
        \cline{2-6}
        \multirow{-7}{*}{\STAB{\rotatebox[origin=c]{90}{{Validation set}}}}{} &  Model$_{multiclass}$  &  96.63 & 0.979  & 0.986  & 0.882\\[0.3em]
        \hline \hline
        {}  & Model$_{2.5x}$ & 99.07 & 0.990 & 0.989 & 0.998 \\
       
        \multirow{-2}{*}{\STAB{\rotatebox[origin=c]{90}{{\shortstack{Test\\set}}}}}  & Model$_{multiclass}$ & 98.27 & 0.992 & 0.990 & 0.973  \\
        \hline
        \end{tabular}}
        \label{table:1} 
\end{table}

\section{Results and Discussion}


\subsection{Patch-wise Classification}
In this experiment, we compare the patch-wise classification performance of binary and multiclass models against the baseline model and literature. Table~\ref{table:1} shows the results of our proposed method on both the validation and test set. Binary models developed on lower magnification levels perform relatively better than other counterparts. Model$_{2.5x}$ outperforms all other models in most evaluation metrics. It significantly outperformed the baseline model with a nearly 5\% F1 score. However, Model$_{10x}$ gives the highest sensitivity value, which shows that context is more important than cellular details at a high level.
Model$_{multiclass}$ resulted in higher accuracy than the baseline model and literature but did not surpass binary models. It might be due to the misclassification of normal tissue as benign lesions. In addition to this, benign and malignant lesions are located close to the epidermis and may contain some normal tissue due to the rough annotations. On the test set, multiclass models exhibited higher F1 and sensitivity in predicting malignant lesions. In contrast, binary models resulted in false positives and predicted normal adnexal structures as benign lesions.

\begin{table}[ht!]
    \centering
      \caption{The results for the slide-level prediction on the test set, using best binary and multiclass models.}
    \resizebox{\columnwidth}{!}{\begin{tabular}{||c| c|c|c||}
        \cline{2-4}
        \multicolumn{1}{c|}{} &  \textbf{Melanoma} & \textbf{Benign Nevus} & \textbf{Total (Acc.)} \\ [0.2em]
        \hline 

        Model$_{2.5x}$  &  4 & 5 & 9 (100 \%)\\
        Model$_{multiclass}$ & 4 & 5 & 9 (100 \%) \\
        \hline
        Ground Truth & 4 & 5 & \multicolumn{1}{c}{} \\[0.1em]
        \cline{1-3} 
        \end{tabular}}
        \label{table:2} 
\end{table}


\subsection{Slide-level Prediction}
We choose the binary Model$_{2.5x}$ and Model$_{multiclass}$ from the previous experiment for slide-level prediction. The values for $t_{p}$ and $t_{r}$ are found to be 0.999 and 0.04, respectively, based on the validation set and fixed for the test. All patients in the test set were accurately predicted as shown in Table~\ref{table:2}. 

Fig.~\ref{fig:local} depicts the localization maps over some example WSIs from the test set. In the first row, the binary model detects some unannotated tissue as a malignant lesion. Besides, the multiclass model closely identifies melanoma, as marked in the annotation. It points to small benign lesions on unannotated areas as well. 
In the second row, the binary model predicts normal adnexal structure as a benign lesion because the model is unaware of normal tissue morphology. The multiclass model predicts some tissue (on the left and right) as benign lesions.   
Overall, the binary model overestimates the size of lesions; however, the multiclass model tightly localizes lesions inside the annotated boundary. Both models successfully predict the presence of melanoma and benign nevus at the slide level. The proposed method is computationally friendly for clinical practice and takes roughly three minutes per WSI to obtain a diagnostic outcome and ROI.

\section{Conclusion and Future Work}
\label{sec:conclusion}
Computational pathology systems may assist pathologists by providing automatic diagnosis and locating regions of interest. In this work, we proposed a CNN-based method to classify patches and create a localization map to detect and segment lesions in whole slide images from skin biopsies and identify patients with melanoma. Interestingly, our method uses a single CNN network to perform lesion segmentation and do slide-level melanoma classification with one extra learned parameter. Models developed on lower magnification levels accurately provide slightly better patch-wise results indicating that context is important. Overall, the proposed method gives promising results and demonstrates its efficacy as a diagnostic service for clinical practices.

In future work, the proposed method should be validated with a larger dataset. The inclusion of tumor necrosis annotations may be beneficial as an additional diagnostic factor, as it is a vital hallmark of rapid cell proliferation.

\section*{Compliance with Ethical Standards}
\label{sec:ethics}
This study was performed in line with the principles of the Declaration of Helsinki. Approval was granted by the ethics committee of the hospital (No\#.2019/747/REK vest). 

\section*{Acknowledgment}
\label{sec:acknowledgments}
This work is financially supported by CLARIFY Project. CLARIFY is a research and innovation program under the Marie Skłodowska-Curie grant agreement No. 860627. The work of Helga Hardardottir is financed through the project “Pathology Services in the Western Norway Health Region – a center for applied digitization” from a Strategic investment from the Western Norway Health Authority.   
The authors have no relevant financial or non-financial interests to disclose.


\bibliographystyle{IEEEtran}
\bibliography{ref}

\end{document}